\documentclass[aps,prc,nofootinbib,a4paper,twocolumn,nofootinbib]{revtex4-2}
\usepackage[utf8]{inputenc}
\usepackage{amsmath}
\usepackage{bm}
\usepackage{graphicx}
\usepackage{color}

\newcommand{\dd}{{\rm d}}
\newcommand{\ee}{{\rm e}}
\newcommand{\ii}{{\rm i}}
\newcommand{\mean}[1]{\left\langle #1 \right\rangle} 

\newcommand{\ffs}{f_{\rm f.s.}}
\newcommand{\Kn}{\mathrm{Kn}}
\newcommand{\vecp}{\!\vec{\,p}}
\newcommand{\vrel}{v_{\rm rel.}}

\begin{document}

\title{Early time behavior of spatial and momentum anisotropies in a kinetic approach to nuclear collisions}

\author{Marc Borrell} \email{marcborrell@physik.uni-bielefeld.de}
\author{Nicolas Borghini} \email{borghini@physik.uni-bielefeld.de}

\affiliation{Fakult\"at f\"ur Physik, Universit\"at Bielefeld, 
D-33615 Bielefeld, Germany}

\date{\today}

\begin{abstract}
We derive a general formula for the early time dependence of a phase space distribution evolving according to the kinetic Boltzmann equation. 
Assuming that the early evolution of the system created in high-energy nuclear collisions can be described by kinetic theory, we calculate the scaling behaviors for the onset of various characteristics of the transverse dynamics. 
In particular, we show that the scaling behavior of the anisotropic flow coefficients $v_n$ at early times does not depend on the details of the collision kernel or the system composition, while at  the same time it differs from the prediction of fluid dynamics.
\end{abstract}

\maketitle

\section{Introduction}
\label{s:intro}

Ultrarelativistic nuclear collisions at the Relativistic Heavy-Ion Collider (RHIC) and the Large Hadron Collider (LHC) produce a large number of particles, whose bulk, consisting of soft hadrons, exhibits many signals of collective dynamical behavior. 
The latter, referred to as collective flow, is almost universally interpreted as the genuine space-time evolution of the created system.
One of the key observables of bulk collectivity involved is anisotropic flow, i.e.\ the asymmetry in the transverse emission pattern of particles~\cite{Heinz:2013th,Bhalerao:2020ulk}. 

The collective dynamical evolution of the bulk is commonly modeled as the dissipative expansion of a relativistic fluid, at least for a significant part of the system history.
Such modeling, possibly supplemented with a dynamical ``pre-equilibrium'' stage (see Refs.~\cite{Schlichting:2019abc,Berges:2020fwq} for recent reviews) and a hadronic afterburner following the fluid-dynamical evolution, yields a very good description of experimentally measured soft hadron production in collisions of heavy nuclei. 
More surprisingly, it can also satisfactory describe the bulk in collisions of smaller systems with a large final-state multiplicity~\cite{Nagle:2018nvi}.

The use of fluid dynamics is however more questionable in such systems with few degrees of freedom~\cite{Schenke:2021mxx}. 
Moreover, since the overall system lifetime is shorter, the pre- and post-equilibrium stages become comparatively more important in the evolution. 
This has led to a renewed interest in alternative descriptions of collective flow, in particular in kinetic-theory models. 
In this framework, a number of recent studies investigated anisotropic flow for setups with a simplified initial geometry~\cite{Romatschke:2018wgi,Kurkela:2018ygx,Borghini:2018xum,Kurkela:2019kip,Kurkela:2020wwb,Kurkela:2021ctp,Ambrus:2021fej} --- to put aside the uncertainties in the initial state of actual small systems ---, revisiting and extending earlier approaches~\cite{Heiselberg:1998es,Alver:2010dn,Borghini:2010hy}.

In this spirit, we consider in this paper a system of degrees of freedom described by a single-particle phase space distribution $f(t,\vec{x},\vecp)$ obeying the kinetic Boltzmann equation~\cite{DeGroot:1980dk} 
\begin{equation}\label{Boltzmann-eq}
p^\mu\partial_\mu f(t,\vec{x},\vecp) = {\cal C}[f],
\end{equation}
with ${\cal C}[f]$ the collision term modeling the effect of rescatterings, whose dependence on the position $\vec{x}$ and momentum $\vecp$ we do not denote. 
As is customary in the study of nuclear collisions at ultrarelativistic energies, we do not include the possible influence of a mean field on the left hand side of this equation.
We focus on the development of anisotropic transverse flow and the system geometry at early times, investigating their respective scaling behaviors as a function of time and of the average number of rescatterings per particle, characterized by an inverse Knudsen number. 

The time evolution of anisotropic flow and spatial characteristics of the system created in heavy-ion collisions has already been studied before, either in fluid-dynamical~\cite{Heinz:2002rs,Kolb:2002cq,Kolb:2003dz,Teaney:2010vd} or transport~\cite{Alver:2010dn,Ambrus:2021fej} calculations, extended over the whole system lifetime, yet always for specific initial conditions. 
In contrast, it was found in Ref.~\cite{Vredevoogd:2008id} that irrespective of the initial state, the early-time development of (the azimuthally symmetric component of) transverse flow seems to be universal, with a  growth linear in time for various classes of models. 
Here we do not investigate several classes of models but ``only'' kinetic theory --- yet at a quite general level --- and we also do not specify the initial state of the system but leave it arbitrary.

We begin in Sect.~\ref{s:free-stream} by discussing the time development of the quantities we are interested in in a collisionless system. 
In Sect.~\ref{s:interacting-system1}, we introduce the early-time expansion of the phase space distribution, following from the Boltzmann equation, which we later use to derive the behavior of anisotropies in an interacting system. 
We then investigate in Sect.~\ref{s:interacting-system2} the special case of a two-dimensional system of massless particles with elastic binary rescatterings, before generalizing in Sect.~\ref{s:generalization} to the case of a generic kinetic theory, before we discuss our results (Sect.~\ref{s:discussion}). 
A few calculations relevant to Sect.~\ref{s:interacting-system1}--\ref{s:generalization} are included in appendices. 
Throughout the paper we set $c = \hbar = 1$ and we use a metric with negative signature.

\section{Free streaming system}
\label{s:free-stream}

Let us first investigate a non-interacting system, for which the collision term on the right hand side of Eq.~\eqref{Boltzmann-eq} vanishes. 
How anisotropic flow and the spatial eccentricities behave in such a system is well known. 
Since there are no rescatterings, the momentum distribution of the particles cannot change, in particular the flow harmonics $v_n$. 
In turn, in the absence of initial anisotropic flow the spatial eccentricities monotonously decrease in absolute value, tending towards 0 at large times. 
We shall nevertheless discuss the case of such a system in some detail, first, to introduce a few notations. 
And secondly, because our results in this section will prove to be useful when we consider the early-time evolution of an interacting system, in which the calculations will be performed in the vicinity of the non-interacting case. 

As is well known, the solutions of the collisionless Boltzmann equation are free-streaming solutions obeying the functional relation
\begin{equation}
\label{free-streaming-condition}
\ffs(t,\vec{x},\vecp) = \ffs\!\bigg(t_0,\vec{x}-\frac{\vecp}{E}(t-t_0),\vecp\bigg),
\end{equation}
where $t_0$ denotes a reference time, in particular the initial time of the evolution. 
Throughout the paper we shall often denote with a subscript~0 the value at $t_0$ of a function of time (and possibly other variables), as e.g.\ $f_0(\vec{x},\vecp) = f(t_0,\vec{x},\vecp)$.
Without loss of generality, we shall take $t_0=0$ in our calculations. 

Let $g(\vec{x},\vecp)$ be a function of the phase space coordinates.
We denote by $\mean{\cdots}_t^{\rm f.s.}$ an average with a free-streaming distribution $\ffs(t,\vec{x},\vecp)$ taken at time $t$:
\begin{equation}
\mean{g(\vec{x},\vecp)}_t^{\rm f.s.} \equiv 
\int\!g(\vec{x},\vecp)_{}\ffs(t,\vec{x},\vecp)\,\dd^3\vec{x}\,\dd^3\vecp,
\end{equation}
where the integral runs over the whole phase space. 
More generally, all averages throughout the paper are performed over phase space and use a particle-number density as weight.

Using the characteristic relation~\eqref{free-streaming-condition} in the integrand, a straightforward change of variables allows one to express $\mean{g(\vec{x},\vecp)}_t^{\rm f.s.}$ in terms of an average at $t_0$:
\begin{equation}
\label{t-average_vs_t0-average}
\mean{g(\vec{x},\vecp)}_t^{\rm f.s.} = 
\mean{g\big(\vec{x}+\vec{v}\,t,\vecp\big)}_0,
\end{equation}
where $\vec{v} \equiv \vecp/E$.
Note that we drop the the superscript f.s.\ when denoting the average in the initial state.
With the help of this identity one can readily derive the time evolution in a non-interacting system of the spatial ``eccentricities''~\cite{Teaney:2010vd,Gardim:2011xv}
\begin{equation}\label{eccentricities}
\epsilon^{\bf x}_{n\,}\ee^{\ii n\Phi_n} \equiv -\frac{\mean{r_\perp^n\ee^{\ii n\theta}}}{\mean{r_\perp^n}},
\end{equation}
where $(r_\perp, \theta)$ are polar coordinates in the transverse plane evaluated in a centered frame, so that the definition is only interesting for $n\geq 2$.
For simplicity, we assume that the initial phase space distribution is isotropic in momentum space at each point $\vec{x}$. 

Setting $\Phi_2=0$ for the moment, we being with the ``ellipticity''
\begin{equation}
\epsilon_2^{\bf x} \equiv \frac{\mean{y^2-x^2}}{\mean{x^2+y^2}},
\label{eps2def}
\end{equation} 
whose behavior in a free-streaming system has already been investigated~\cite{Kolb:2000sd}. 
If each average in Eq.~\eqref{eps2def} is computed with a free-streaming solution, relation~\eqref{t-average_vs_t0-average} yields 
\[
\mean{x^2}_t^{\rm f.s.} = \mean{(x+v_xt)^2}_0 = 
\mean{x^2}_0 + t^2\mean{v_x^2}_0,
\]
where we used $\mean{v_x}_0=0$ as follows from the assumed isotropy in momentum space. 
Similarly, one has
\[
\mean{y^2}_t^{\rm f.s.} = \mean{y^2}_0 + t^2\mean{v_y^2}_0. 
\]
Momentum-space isotropy also yields 
\[
\mean{v_x^2}_0 = \mean{v_y^2}_0 = \frac{1}{2}\mean{v_\perp^2}_0,
\]
where $v_\perp = |\bm{v}_\perp|$ is the modulus of the transverse velocity, so that one quickly finds~\cite{Kolb:2000sd}
\begin{equation}
\label{fs-eps2^x(t)}
\epsilon_2^{\bf x}(t) = 
\frac{\epsilon_2^{\bf x}(t_0)}{1 + \mean{v_\perp^2}_0t^2/\mean{r_\perp^2}_0},
\end{equation}
where $\mean{r_\perp^2}_0\equiv\mean{x^2+y^2}_0$.
This result is unchanged if the initial participant-plane angle $\Phi_2$ is not aligned with the $x$-axis: 
in a non-interacting system without initial flow, $\Phi_2$ does not change with time. 
As anticipated, (the modulus of) $\epsilon_2^{\bf x}(t)$ decreases with time, with an early-time departure from its initial value that is quadratic in time:
\begin{equation}
\label{fs-eps2^x(t)_small-t}
\epsilon_2^{\bf x}(t) \simeq 
\epsilon_2^{\bf x}(t_0)\bigg( 1 - \frac{\mean{v_\perp^2}_0}{\mean{r_\perp^2}_0}_{}t^2\bigg)
\ \text{for }t\ll\Bigg(\!\frac{\mean{r_\perp^2}_0}{\mean{v_\perp^2}_0}\!\Bigg)^{\!\!1/2}.
\end{equation}  
As we shall now show, the spatial eccentricity $\epsilon_n^{\bf x}$ in the $n$-th harmonic also departs quadratically with time from its initial value in a free streaming system without initial anisotropic flow.\footnote{To the best of our knowledge, this behavior has not been reported in the literature before.}
Consider thus
\begin{equation}
\label{fs-eps_n_calc0}
\epsilon^{\bf x}_{n\,}\ee^{\ii n\Phi_n} = -\frac{\mean{r_\perp^n\ee^{\ii n\theta}}}{\mean{r_\perp^n}} = 
-\frac{\mean{(x + \ii_{}y)^n}}{\mean{(x^2 + y^2)^{n/2}}}.
\end{equation} 
In the case of a non-interacting system, invoking Eq.~\eqref{t-average_vs_t0-average} for the numerator yields
\begin{align*}
\mean{(x + \ii_{}y)^n}_t^{\rm f.s.} &= \mean{\big[(x+v_xt)+\ii(y+v_yt)\big]^n}_0 \\
&= \mean{\big[(x+\ii_{}y)+(v_x+\ii_{}v_y)t\big]^n}_0.
\end{align*}
Introducing the azimuthal angle $\varphi$ of the velocity (or equivalently the momentum), we can switch back to polar coordinates: $x+\ii_{}y = r_\perp\ee^{\ii_{}\theta}$ and 
$v_x+\ii_{}v_y = v_\perp\ee^{\ii_{}\varphi}$.
This yields
\[
\mean{r_\perp^n\ee^{\ii n\theta}}_t^{\rm f.s.} = 
\sum_{k=0}^n\binom{n}{k}\!\mean{r_\perp^{n-k}\ee^{\ii(n-k)\theta\,}
v_\perp^k\ee^{\ii_{}k\varphi}}_{\!0}t^k.
\]
Due to the local isotropy in momentum space, every average of $\ee^{\ii_{}k\varphi}$ with $k\neq 0$ vanishes, so that the above average is actually time-independent:
\begin{equation}
\label{<r^nexp(intheta)>_fs}
\mean{r_\perp^n\ee^{\ii n\theta}}_t^{\rm f.s.} = \mean{r_\perp^n\ee^{\ii n\theta}}_0.
\end{equation}
In particular the phase of $\mean{r_\perp^n\ee^{\ii n\theta}}_t^{\rm f.s.}$ remains constant, i.e.\ the participant-plane angle $\Phi_n$ does not rotate in the absence of interactions, which is quite intuitive. 

For the denominator of Eq.~\eqref{fs-eps_n_calc0}, we similarly write
\begin{align}
\mean{r_\perp^n}_t^{\rm f.s.} = \mean{\big[(x+v_xt)^2 + (y+v_yt)^2\big]^{n/2}}_0.
\end{align}
If $n$ is even, the quantity to be averaged can be exactly computed for any $t$ with the binomial theorem. 
Invoking the isotropy in momentum space, only the terms with even powers of both $v_x$ and $v_y$ remain after averaging. 
If we only look at the early-time behavior, a Taylor expansion valid for both even and odd values of $n$ yields
\begin{align*}
\mean{\big[(x+v_xt)^2 + (y+v_yt)^2\big]^{n/2}}_{\!0} &\simeq \\
\mean{(x^2+y^2)^{n/2}}_{\!0}\bigg[1\,+\,&{\cal O}\bigg(\frac{\mean{\bm{v}_\perp^2}_0}{\mean{r_\perp^2}_0}t^2\bigg)\bigg],
\end{align*}
where the factor multiplying the term in $t^2$ depends on $n$ and is positive. 
All in all, one finds
\begin{equation}
\label{fs-eps_n^x(t)_small-t}
\epsilon_n^{\bf x}(t) \simeq 
\epsilon_n^{\bf x}(t_0)\bigg[ 1 - {\cal O}\bigg(\frac{\mean{v_\perp^2}_0}{\mean{r_\perp^2}_0}_{}t^2\bigg)\bigg]
\ 
\text{for }t\ll\!\Bigg(\!\frac{\mean{r_\perp^2}_0}{\mean{v_\perp^2}_0}\!\Bigg)^{\!\!1/2},
\end{equation}  
similar to Eq.~\eqref{fs-eps2^x(t)_small-t}.

The results of this section can be extended in a straightforward way to the generalized spatial eccentricities~\cite{Teaney:2010vd,Gardim:2011xv} 
\begin{equation}
\label{eps_n,m_def}
\epsilon^{\bf x}_{n,m\,}\ee^{\ii n\Phi_{m,n}} \equiv -\frac{\mean{r_\perp^m\ee^{\ii n\theta}}}{\mean{r_\perp^m}}.
\end{equation}
In a free-streaming system without initial anisotropic flow, the numerator of the ratio on the right hand side is actually independent of time. 
In turn, the denominator increases with $t^2$, so that eventually one finds the same behavior as in Eq.~\eqref{fs-eps_n^x(t)_small-t}, with a different factor in front of the term in $t^2$.

\section{Interacting system: general idea}
\label{s:interacting-system1}

In this section we introduce a general early-time expansion of the phase space distribution of an interacting system obeying the Boltzmann equation, by carefully exploiting the latter. 
This expansion will form the basis of our calculations of the development of anisotropic flow and spatial eccentricities in the following two sections. 

Consider an arbitrary single-particle phase space distribution $f(t,\vec{x},\vecp)$. 
(Strictly speaking, the distribution needs to vanish quickly enough as $|\vec{x}|$ or $|\vecp|$ go to infinity to be normalized to the number of particles in the system. In addition, we assume that it is sufficiently continuously differentiable for our equations to make sense.)
If one is interested in the early-time evolution starting from a known initial condition at $t=t_0$, one can begin with the Taylor expansion
\begin{align}
f(t,\vec{x},\vecp) = f_0(\vec{x},\vecp) &+ 
t\, \partial_t f(\vec{x},\vecp)\big|_0 \cr 
& + \frac{t^2}{2} \partial_t^2 f(\vec{x},\vecp)\big|_0  + \cdots
\label{Taylordg}
\end{align} 
where $f$ and its successive time derivatives are evaluated at the initial time $t_0=0$.
The early-time behavior is thus governed by these time derivatives.

Making use of the relativistic Boltzmann equation~\eqref{Boltzmann-eq}, the first time derivative in Eq.~\eqref{Taylordg} trivially reads
\begin{equation}
\partial_t f(\vec{x},\vecp)\big|_0 = 
-\frac{\vecp}{E}\cdot\vec{\nabla}_{\!x} f(\vec{x},\vecp)\big|_0 + 
\frac{1}{E}_{}{\cal C}[f]\big|_0.
\label{be}
\end{equation} 
If we now differentiate the Boltzmann equation with respect to time, we obtain an expression for $\partial_t^2f$ as a function of $\partial_t\vec{\nabla}_{\!x}f$ and the time derivative $\partial_t{\cal C}[f]$ of the collision term.
Exchanging the order of time derivative and gradient in the term $\partial_t\vec{\nabla}_{\!x}f = \vec{\nabla}_{\!x}\partial_t f$ and replacing $\partial_t f$ by its expression as given by Eq.~\eqref{Boltzmann-eq}, one finds
\begin{align}
\partial_t^2 f(t,\vec{x},\vecp) = 
& \,\bigg(\frac{\vecp}{E}\cdot\vec{\nabla}_{\!x}\bigg)^{\!\!2} f(t,\vec{x},\vecp) \cr
& -\frac{\vecp}{E^2}\cdot\vec{\nabla}_{\!x}{\cal C}[f] +
\frac{1}{E}_{}\partial_t{\cal C}[f],
\end{align}
which can then be evaluated at $t_0$.
Given an explicit expression for the collision term ${\cal C}[f]$, the time derivative $\partial_t{\cal C}[f]|_0$ will generically involve $\partial_t f|_0$, which can again be replaced by the right hand side of Eq.~\eqref{be}.
We can thus eliminate time derivatives at $t_0$ in a systematic manner, replacing them by expressions that only involve the spatial dependence of the initial distribution $f_0(\vec{x},\vecp)$.  

Iterating this approach, one finds 
\begin{widetext}
\begin{align}
f(t,\vec{x},\vecp) = f_0(\vec{x},\vecp) & +
t\bigg(\!\!-\!\frac{\vecp}{E}\cdot\vec{\nabla}_{\!x} f(\vec{x},\vecp) + 
\frac{1}{E}_{}{\cal C}[f] \bigg)_{\!0} +
\frac{t^2}{2} \bigg(\frac{\big(\vecp\cdot\vec{\nabla}_{\!x}\big)^{\!2}}{E^2} f(\vec{x},\vecp) -
\frac{\vecp}{E^2}\cdot\vec{\nabla}_{\!x} {\cal C}[f] + 
\frac{1}{E}_{}\partial_t {\cal C}[f] \bigg)_{\!0} \cr
& + \frac{t^3}{3!}\bigg(\!\!-\!\frac{\big(\vecp\cdot\vec{\nabla}_{\!x}\big)^{\!3}}{E^3} f(\vec{x},\vecp) + 
\frac{\big(\vecp\cdot\vec{\nabla}_{\!x}\big)^{\!2}}{E^3} {\cal C}[f]  - 
\frac{\vecp}{E^2}\cdot\vec{\nabla}_{\!x} \partial_t {\cal C}[f] +
\frac{1}{E}_{}\partial_t^2 {\cal C}[f] \bigg)_{\!0} \cr
& +\frac{t^4}{4!}\bigg(\frac{\big(\vecp\cdot\vec{\nabla}_{\!x}\big)^{\!4}}{E^4} f(\vec{x},\vecp) - 
\frac{\big(\vecp\cdot\vec{\nabla}_{\!x}\big)^{\!3}}{E^4} {\cal C}[f] + 
\frac{\big(\vecp\cdot\vec{\nabla}_{\!x}\big)^{\!2}}{E^3} \partial_t {\cal C}[f] - 
\frac{\vecp}{E^2}\cdot\vec{\nabla}_{\!x} \partial_t^2 {\cal C}[f] +
\frac{1}{E}_{}\partial_t^3 {\cal C}[f]\bigg)_{\!0} \cr
&+ {\cal O}(t^5).
\label{Taylordg2}
\end{align}
This lengthy expression can be shortened if one realizes that the terms that do not involve the collision kernel ${\cal C}[f]$ or its derivatives are actually the successive time derivatives of the free-streaming distribution $\ffs(t,\vec{x},\vecp)$ that coincides with $f_0(\vec{x},\vecp)$ at the initial time $t_0$, see Eq.~\eqref{free-streaming-condition}.
Thus we may rewrite Eq.~\eqref{Taylordg2} as
\begin{align}
f(t,\vec{x},\vecp) = \ffs(t,\vec{x},\vecp) &+ t\,\frac{{\cal C}[f]\big|_0}{E}  + 
\frac{t^2}{2} \bigg(\!\!-\!\frac{\vecp}{E^2}\cdot\vec{\nabla}_{\!x} {\cal C}[f] +
\frac{1}{E}_{}\partial_t {\cal C}[f]\bigg)_{\!0} \cr
& +\frac{t^3}{3!}\bigg( \frac{\big(\vecp \cdot \vec{\nabla}_{\!x}\big)^{\!2}}{E^3} {\cal C}[f]  - 
\frac{\vecp}{E^2}\cdot\vec{\nabla}_{\!x} \partial_t {\cal C}[f] +
\frac{1}{E}_{}\partial_t^2 {\cal C}[f] \bigg)_{\!0} \cr
& +\frac{t^4}{4!}\bigg(\!\!-\!
\frac{\big(\vecp\cdot\vec{\nabla}_{\!x}\big)^{\!3}}{E^4} {\cal C}[f] + 
\frac{\big(\vecp\cdot\vec{\nabla}_{\!x}\big)^{\!2}}{E^3} \partial_t {\cal C}[f] - 
\frac{\vecp}{E^2}\cdot\vec{\nabla}_{\!x} \partial_t^2 {\cal C}[f] +
\frac{1}{E}_{}\partial_t^3 {\cal C}[f]\bigg)_{\!0} + {\cal O}(t^5).\ \quad
\label{Taylordg3}
\end{align} 
\end{widetext}
Note that the first term on the right hand side actually resums all orders in $t$, i.e.\ is automatically valid at any order in $t$ in the absence of interactions. 
In the following sections we shall exploit Eq.~\eqref{Taylordg3} to investigate how the early time evolution of geometric eccentricities or anisotropic flow coefficients departs in the presence of rescatterings from their free-streaming behavior.

Before that, let us further discuss expansion~\eqref{Taylordg3}. 
The coefficient of the linear term in $t$ and the first terms in the factors within parentheses multiplying the higher-order powers of $t$ are of the generic form
\begin{equation}
\label{terms_O(d_t^0C)}
\frac{\big(\!\!-\!\vecp \cdot \vec{\nabla}_{\!x}\big)^k}{E^{k+1}} {\cal C}[f]\big|_0,
\end{equation}
with $k\geq 0$.
These contributions only involve the initial phase space distribution $f_0$ and its spatial derivatives, which can be signaled by writing the collision term ${\cal C}[f_0]$. 
If $\sigma$ is a typical cross section for the rescatterings modeled by the collision term, the terms~\eqref{terms_O(d_t^0C)} are of order ${\cal O}(\sigma)$. 
Equivalently, these terms are of order ${\cal O}(\Kn^{-1})$, where $\Kn$ denotes a characteristic Knudsen number built from the mean free path and a typical length scale of the initial state distribution.
Although the definition of $\Kn$ is arbitrary, we shall from now on systematically use ${\cal O}(\Kn^{-1})$ instead of ${\cal O}(\sigma)$, since the Knudsen number is dimensionless while the cross section is not.
Since $\Kn^{-1}$ is roughly speaking a measure of the average number of rescatterings undergone by each particle in the system, it is similar to the opacity used in a number of related studies~\cite{Kurkela:2018ygx,Kurkela:2019kip,Kurkela:2020wwb,Ambrus:2021fej}.

Starting from order $t^2$ the expansion~\eqref{Taylordg3} contains terms of the form
\begin{equation}
\label{terms_O(d_t^1C)}
\frac{\big(\!\!-\!\vecp \cdot \vec{\nabla}_{\!x}\big)^k}{E^{k+1}} \partial_t{\cal C}[f]\big|_0
\frac{t^{k+2}}{(k+2)!},
\end{equation} 
with $k\geq 0$. 
As was already mentioned, the time derivative $\partial_t{\cal C}[f]$ can be computed when the collision kernel is known, by replacing every $\partial_t f$ as in Eq.~\eqref{be}.
Accordingly, the term~\eqref{terms_O(d_t^1C)} for a given $k$ will yield two types of contributions: 
on the one hand, terms involving $(k+1)$-th spatial derivatives of $f_0$ while being of order ${\cal O}(\Kn^{-1})$, like the contributions of the form~\eqref{terms_O(d_t^0C)}. 
On the other hand, there also come terms of order ${\cal O}(\Kn^{-2})$, with spatial derivatives of $f_0$ of order $k$ only. 
As an example, we present in Appendix~\ref{app:d_tC[f]} the calculation of $\partial_t{\cal C}[f]$ for a specific choice of collision kernel.

More generally, transforming the factor in front of the term of order $t^k$ with $k\geq 1$ in Eq.~\eqref{Taylordg3} so as to express every time derivative at $t_0$ in terms of spatial gradients of $f_0$ --- possibly entering (iterated) collision kernels ---, one finds that the factor involves contributions at order $\Kn^{-1}$, $\Kn^{-2}$, \ldots $(\Kn^{-1})^k$. 
Equation~\eqref{Taylordg3} thus implicitly contains a double expansion in powers of both $t$ and $\Kn^{-1}$, where for consistency only the powers $(\Kn^{-1})^j$ with $j\leq k$ should appear at order $t^k$.
Conversely, corrections to the free-streaming distribution of order $(\Kn^{-1})^j$ only come up at order $t^j$ and higher in Eq.~\eqref{Taylordg3}.
Thus, we are in principle able to go beyond the linear order in $\Kn^{-1}$ to which the (semi-)analytical results existing in the literature~\cite{Heiselberg:1998es,Borghini:2010hy,Borghini:2018xum,Kersting:2018qvi,Ambrus:2021fej} are usually restricted.

\section{Early-time behavior of a two-dimensional system of massless particles with elastic rescatterings}
\label{s:interacting-system2}

To illustrate how the ideas introduced in the previous section can be used to derive the early time behavior of quantities characterizing the expanding system, we shall now choose a specific ansatz for the collision kernel ${\cal C}[f]$ of the Boltzmann equation. 
Since it is possibly the simplest possible case, we consider a system of massless particles without spin, undergoing elastic binary collisions, and propagating in two dimensions\footnote{Two-dimensional vectors will be denoted in boldface: ${\bf x}$, ${\bf p}$.} only, namely the transverse plane of a nucleus-nucleus collision. 
As we discuss below, the latter assumption --- which will be relaxed in Sect.~\ref{ss:3D} --- together with that of massless particles significantly simplifies the form of the M{\o}ller velocity\footnote{We use the notation $\vrel$, although the M{\o}ller velocity does not coincide with the relative velocity.}
\begin{equation}
\label{vrel}
\vrel = \sqrt{(\vec{v}-\vec{v}_1)^2-|\vec{v}\times\vec{v}_1|^2} 
\end{equation}
in the collision integral~\cite{DeGroot:1980dk}
\begin{equation}
\label{C[f]_2->2}
{\cal C}[f] = \frac{E}{2}\!\int\!(f'f'_1 - ff_1)_{}\vrel
\frac{\dd\sigma}{\dd\Theta}\,
\dd\Theta\,\dd^2{\bf p}_1,
\end{equation} 
where we did not denote the arguments of the phase space distribution before (unprimed) or after (primed) a collision, while $\Theta$ is the scattering angle and $\dd\sigma/\dd\Theta$ the corresponding differential cross section.
To remain as general as possible, we do not specify the latter --- which in the two-dimensional case has the dimension of a length ---, nor the form of the initial phase space distribution. 
Instead of Eq.~\eqref{C[f]_2->2}, we shall also use the equivalent form~\cite{DeGroot:1980dk}
\begin{align}
\label{C[f]_2->2_sym}
{\cal C}[f] = \frac{1}{2}&\!\int\!(f'f'_1 - ff_1)_{}
W({\bf p},{\bf p}_1\to{\bf p}',{\bf p}'_1)\cr
&\hspace{2cm}\times\frac{\dd^2{\bf p}_1}{(2\pi)^2E_1}\frac{\dd^2{\bf p}'}{(2\pi)^2E'}
\frac{\dd^2{\bf p}'_1}{(2\pi)^2E'_1},\qquad
\end{align}
with the transition rate $W({\bf p},{\bf p}_1\to{\bf p}',{\bf p}'_1)$, which has the advantage to be more easily generalized to other types of rescatterings. 

We shall assume that the initial distribution $f_0({\bf x},{\bf p})$ is isotropic in momentum space at each point ${\bf x}$, so that there is no initial anisotropic flow in the system. 
In contrast, the spatial distribution in the initial state is asymmetric and in particular depends on the polar angle $\theta$.
To represent this variation with $\theta$, we symbolically write 
\begin{equation}
\label{f0_vs_theta}
f_0 = \bar{f}_0 + \sum_{n\neq 0}\delta_n f_0^{(n)}\ee^{-\ii_{}n\theta},
\end{equation}
where the real-valued functions $\bar{f}_0$ and $f_0^{(n)}$ are independent of $\theta$, while $\delta_n$ is a dimensionless complex number characterizing the modulation of period $2\pi/n$ in $\theta$. 
To ensure that $f_0$ is real-valued, both $\delta_{-n} = \delta_n^*$ and $f_0^{(-n)}=f_0^{(n)}$ should hold for every $n$.
Clearly, the eccentricity $\epsilon_n^{\bf x}$ and its generalizations~\eqref{eps_n,m_def} will be proportional to $\delta_n$:
\begin{equation}
\epsilon_{n,m}^{\bf x} \propto \delta_n.
\end{equation}
Equation~\eqref{f0_vs_theta} is meant to be schematic and to stand for the polar dependence of any systematic expansion relying on a true complete basis of functions on the transverse plane, like the cumulant~\cite{Teaney:2010vd} or the Bessel--Fourier~\cite{ColemanSmith:2012ka,Floerchinger:2013vua} expansions.

\subsection{Anisotropic flow}
\label{ss:vn}

In a kinetic description, anisotropic flow results from the rescatterings in the system, which convert the asymmetry of its initial geometry into an an anisotropy in momentum space.
The Fourier coefficients quantifying the momentum-distribution anisotropies can be obtained from the phase space distribution:
\begin{equation}
\label{vn_vs_f(t,x,p)}
v_n(t) \equiv \big|v_n(t)\big|_{}{\rm e}^{{\rm i}_{}n\Psi_n(t)} = 
\frac{\displaystyle\int\!f(t,{\bf x},{\bf p})\,{\rm e}^{{\rm i}_{}n\varphi_{\bf p}}\,{\rm d}^2{\bf x}\,{\rm d}^2{\bf p}}%
{\displaystyle\int\!f(t,{\bf x},{\bf p})\,{\rm d}^2{\bf x}\,{\rm d}^2{\bf p}},
\end{equation}
where $\varphi_{\bf p}$ denotes the azimuth of ${\bf p}$.
While the integral over ${\bf x}$ runs over the whole position space, that over ${\bf p}$ can either be over the whole momentum space or restricted to an interval in $|{\bf p}|$ (while still running over the whole range for $\varphi_{\bf p}$). 
In the following we solely discuss the first possibility (``integrated flow''), since the second one depends more crucially on the details of the microscopic interaction rate and of the initial phase space distribution.
Accordingly, we do not consider directed flow $v_1$, since its integrated value is fixed to its initial value (here zero) by global momentum conservation.

When the integration runs over the whole phase space, the denominator of Eq.~\eqref{vn_vs_f(t,x,p)} is simply the total number of particles in the system. 
Since we assume in this section that the latter only has elastic scatterings, this number remains constant in time.

\subsubsection{Leading contribution at early times}
\label{sss:vn_O(Kn^-1)}

Since the denominator of Eq.~\eqref{vn_vs_f(t,x,p)} is constant, the whole time dependence of $v_n(t)$ comes from the numerator, in which we can now substitute the early-time expansion~\eqref{Taylordg3} for $f(t,{\bf x},{\bf p})$.
The first term from the free-streaming distribution $\ffs$ does not contribute to $v_n(t)$, since we assumed that there is no initial anisotropic flow. 
The next term linear in $t$ does not contribute either: it consists of the collision kernel ${\cal C}[f_0]$ in the initial state, and every phase space distribution entering it is isotropic in momentum space.\footnote{To be more precise, a dependence on $\varphi_{\bf p}$ appears in ${\cal C}[f_0]$ via the relative velocity. But it appears in the form $\varphi_{\bf p}-\varphi_1$ relative to the azimuth $\varphi_1$ of the ``collision partner'' labeled with 1, and thus disappears in the integration over ${\bf p}_1$ with a momentum-isotropic distribution $f_1\big|_0$.} 

The term quadratic in time in expansion~\eqref{Taylordg3} is the first one involving a contribution of the form~\eqref{terms_O(d_t^0C)} with $k\geq 1$, namely $-({\bf p}/E^2)\cdot\bm{\nabla}_{\!x}{\cal C}[f]\big|_0$.
This term does not contribute to anisotropic flow at early times: 
performing first the integration over the transverse plane in Eq.~\eqref{vn_vs_f(t,x,p)}, one encounters the integral
\begin{equation}
\label{int(d_xC[f])=0}
\int\!\bm{\nabla}_{\!x}{\cal C}[f]\big|_0\,\dd^2{\bf x} = 0,
\end{equation}
since the collision kernel vanishes at infinity.
More generally, at every order in expansion~\eqref{Taylordg3} the terms involving (powers of) the spatial gradient of ${\cal C}[f]\big|_0$ or of its time derivatives will not contribute to the early-time behavior of anisotropic flow, thanks to a similar argument: 
the integration over the transverse plane yields zero at once.
Thus the only terms in Eq.~\eqref{Taylordg3} that can eventually contribute to $v_n(t)$ at early times are those of the form $(1/E)\partial_t^k{\cal C}[f]\big|_0$, without spatial gradient in front, where the $k$-th time derivative of the collision kernel appears at order $t^{k+1}$.

Although $(1/E)\partial_t^k{\cal C}[f]\big|_0$ is not a total spatial gradient, still it contains (powers of) $\bm{\nabla}_{\!x}$ --- multiplied by a momentum --- when one computes explicitly the time derivative, as exemplified in Eqs.~\eqref{d_tC[f]_example}--\eqref{d_t^2C[f]_example}.
Such terms thus couple the spatial and momentum ``components'' of the phase space distribution, via the inner products ${\bf p}\cdot\bm{\nabla}_{\!x}$, under the influence of the particle rescatterings encoded in the transition rate, which is why they are crucial for the development of anisotropic flow. 
Let us sketch how a contribution to the $n$-th harmonic $v_n$ can arise in the explicit calculation of Eq.~\eqref{vn_vs_f(t,x,p)}. 

First, note that ${\rm e}^{{\rm i}_{}n\varphi_{\bf p}}$, or equivalently its real and imaginary parts, can be expressed in terms of powers of the components of ${\bf p}$ and in fact involves the $n$-th power.\footnote{This is of course obvious since ${\rm e}^{{\rm i}_{}n\varphi_{\bf p}} = {\bf p}^n / |{\bf p}|^n$, where we identify the two-dimensional momentum ${\bf p}$ with a complex number $p_x + \ii_{}p_y$.}
Accordingly, any non-zero contribution to $v_n(t)$ must involve a term in ${\bf p}^n$ from the early-time expansion of $f(t,{\bf x},{\bf p})$.
In the time derivative $\partial_t^k{\cal C}[f]\big|_0$, such terms trivially appear if $k=n$, i.e.\ at order $t^{n+1}$. 
Closer investigation reveals that there are already terms in ${\bf p}^n$ at order $t^n$ --- but they do {\em not} contribute to $v_n(t)$ at early times. 
Keeping the discussion to first order in the inverse Knudsen number, these terms come from combining an ``obvious'' $({\bf p}\cdot\bm{\nabla}_{\!x})^{n-1}$ in $\partial_t^{n-1}{\cal C}[f]\big|_0$ with the factor ${\bf p}$ hidden in the M{\o}ller velocity~\eqref{vrel}, which for massless particles reads
\begin{equation}
\label{vrel_2D_m=0}
\vrel = 1-\frac{{\bf p}\cdot{\bf p}_1}{E\,E_1} =
1-\cos(\varphi_{\bf p}-\varphi_1),
\end{equation}
where the second identity only holds in two dimensions (or if the two particles have the same polar angle along the third direction).
As this expression shows, the term that could contribute to $v_n(t)$ at order $t^n$ also involves a multiplicative factor ${\bf p}_1$, i.e.\ will yield an odd function of ${\bf p}_1$ in the integrand of the collision kernel: 
this vanishes in the integration over ${\bf p}_1$, thus giving no contribution to $v_n(t)$.%
\footnote{To be thorough, the term in ${\bf p}\cdot{\bf p}_1$ from the M{\o}ller velocity can contribute to the early time development of $v_n(t)$, but it has to multiply not only a term in $({\bf p}\cdot\bm{\nabla}_{\!x})^{n-1}$, but also a term ${\bf p}_1\cdot\bm{\nabla}_{\!x}$ (or an odd power thereof): 
such a contribution can only appear at order $t^{n+1}$ or higher.}

We mentioned above that the terms from expansion~\eqref{Taylordg3} of the form $(1/E)\partial_t^k{\cal C}[f]\big|_0$ involve contributions of higher order in the inverse Knudsen number $\Kn^{-1}$. 
Careful accounting based on iterating Eq.~\eqref{be} shows that at order $t^{n+1}$ a term in $\big(\Kn^{-1}\big)^j$ is accompanied by $n+1-j$ powers of $\bm{\nabla}_{\!x}$ (see Appendix~\ref{app:d_tC[f]}):
if $j>1$, this cannot contribute to $v_n(t)$. 
Thus the higher orders in $\Kn^{-1}$ affect the early time development of $v_n$ at a higher order in $t$, as we shall discuss in next subsection. 

From the above discussion, any non-zero contribution to $v_n(t)$ at early times actually comes from a term in expansion~\eqref{Taylordg3} that contains the $n$-th power of the spatial gradient $\bm{\nabla}_{\!x}$. 
By investigating explicit examples, or more formally by following a similar approach to that developed in Ref.~\cite{Teaney:2010vd}, one finds that this operator $(\bm{\nabla}_{\!x})^n$ together with the subsequent integration over ${\bf x}$ isolates the component with periodicity $2\pi/n$ in the polar angle $\theta$ of the function it acts upon.
In the term in $\partial_t^n{\cal C}[f]\big|_0$, these gradients appear in the form of products [see Eqs.~\eqref{d_tC[f]_example}--\eqref{d_t^2C[f]_example}]
\[
\bigg[\frac{{\bf p}\cdot\bm{\nabla}_{\!x} f_0({\bf x},{\bf p})}{E}\bigg]^k
\bigg[\frac{{\bf p}_1\cdot\bm{\nabla}_{\!x} f_0({\bf x},{\bf p}_1)}{E_1}\bigg]^{n-k},
\]
with $0\leq k\leq n$.
Replacing each $f_0$ by its schematic expansion~\eqref{f0_vs_theta}, one finds that two kinds of terms contribute to the $n$-th polar mode of $\partial_t^n{\cal C}[f]\big|_0$: 
First, terms involving the symmetric part $\bar{f}_0$ of one factor with the term in $\delta_n f_0^{(n)}$ from the other factor, yielding a contribution proportional to $\delta_n$.
And secondly, terms of the form $\delta_k\delta_{n-k}f_0^{(k)}f_0^{(n-k)}$ with $|k|\geq 1$.
Since every $\delta_k$ is proportional to the eccentricity $\epsilon_k^{\bf x}$, we recover the known fact~\cite{Borghini:2005kd,Teaney:2012ke,Niemi:2012aj,Plumari:2015cfa} that $v_n$ gets a linear contribution in $\epsilon_n^{\bf x}$ together with nonlinear contributions $\epsilon_{n-k}^{\bf x}\epsilon_k^{\bf x}$. 
What we show here for the first time is that in a kinetic-transport approach all contributions to $v_n(t)$ grow with the same power of $t$ at early times, namely $t^{n+1}$:
\begin{equation}
\label{vn_vs_t^(n+1)}
v_n(t) \!\underset{\text{early }t}{\sim}\! 
\Kn^{-1} \bigg(\! {\cal K}^{(1)}_{n,n}\epsilon_n^{\bf x} + 
\sum_{k\geq 1}\!{\cal K}^{(1)}_{n,n-k,k}\epsilon_{n-k}^{\bf x}\epsilon_k^{\bf x}\bigg)
t^{n+1},
\end{equation} 
where the scale defining the notion of an early time is given by the typical transverse size of the system.
The overall growth $v_n(t) \propto t^{n+1}$ in transport computations was already observed in Ref.~\cite{Alver:2010dn,Borghini:2010hy} (and in Refs.~\cite{Heiselberg:1998es,Gombeaud:2007ub} for $v_2$), where it was noted that it differs from fluid-dynamical calculations, which yield the faster increase $v_n(t) \propto t^n$.

Anticipating on the following subsection, let us list which expected contributions to $v_n$ are missing from the scaling behavior~\eqref{vn_vs_t^(n+1)} but appear at higher orders in $t$. 
First, we have already mentioned the terms of higher order in the inverse Knudsen number. 
Secondly, fluid-dynamical simulations~\cite{Noronha-Hostler:2015dbi,Niemi:2015qia}, numerical solution of the Boltzmann equation~\cite{Kurkela:2020wwb} or transport calculations~\cite{Roch:2020zdl} have revealed cubic contributions in the eccentricities to the flow harmonics, as e.g.\ contributions in $(\epsilon_2^{\bf x})^3$ to $v_2$ or $v_6$. 
Such terms are also absent from Eq.~\eqref{vn_vs_t^(n+1)}, because in a scenario with only binary rescatterings and no quantum effects they appear at a higher order in $\Kn^{-1}$~\cite{Borghini:2018xum}.

\subsubsection{Higher order contributions at early times}
\label{sss:vn_O(Kn^-2)}

From the previous subsection we know that at early times anisotropic flow harmonics are ``created'' by the terms of the type $(1/E)\partial_t^k{\cal C}[f]\big|_0$ in expansion~\eqref{Taylordg3}.
For $v_n(t)$, this gives non-zero contributions starting at $k=n$, resulting in the scaling behavior~\eqref{vn_vs_t^(n+1)} in $t^{n+1}$. 
We have also emphasized that a necessary ingredient is the presence of (at least) $n$ powers of the spatial gradient $\bm{\nabla}_{\!x}$ in the expression of $\partial_t^k{\cal C}[f]\big|_0$.

What happens if we push expansion~\eqref{Taylordg3} to order $t^{n+2}$, thus including the term in $\partial_t^{n+1}{\cal C}[f]\big|_0$?
First, the terms of order $\Kn^{-1}$ contain $n+1$ powers of $\bm{\nabla}_{\!x}$, which necessarily multiply $n+1$ momentum variables (${\bf p}$, ${\bf p}_1$\ldots). 
In addition, the contribution may include extra momentum factors --- for example from the ${\bf p}\cdot{\bf p}_1$ term in the M{\o}ller velocity~\eqref{vrel_2D_m=0} ---, which only come in pairs.
Because the momentum variables come in a number of parity opposite to that of $n$, such terms cannot contribute to $v_n(t)$: 
when integrating over all momenta to obtain the integrated $v_n$, the integrand will turn out to be odd in at least one of the momentum variables, and thus yield zero. 
On the other hand, there can be non-vanishing contributions to $v_n(t)$ of order $\Kn^{-1}$ at order $t^{n+3}$, $t^{n+5}$, and so on. 

The derivative $\partial_t^{n+1}{\cal C}[f]\big|_0$ also includes terms of order $\Kn^{-2}$, see for instance the second line of Eq.~\eqref{d_tC[f]_example} in the case $n=0$ or Eq.~\eqref{d_t^2C[f]_example2} for $n=1$.
These terms involve $n$ powers of $\bm{\nabla}_{\!x}$, multiplied with their respective momentum variables.
Thus these terms can yield non-zero contributions to $v_n(t)$ at early times, scaling as $t^{n+2}$.
Moreover, these terms include products of (powers of gradients of) $f$ with (powers of gradients of) the collision kernel ${\cal C}[f]$, and thus are ``cubic'' in the phase space distribution for the Boltzmann kernel~\eqref{C[f]_2->2}--\eqref{C[f]_2->2_sym}.
Accordingly, one quickly sees that the contributions to $v_n(t)$ at order $\Kn^{-2}$ will not only include linear and quadratic terms in the initial anisotropies $\epsilon_n^{\bf x}$, similar to those in the parentheses of Eq.~\eqref{vn_vs_t^(n+1)}, but also cubic terms of the type
\begin{equation}
\label{vn_vs_t^(n+2)}
\sum_{k,l\geq 1}{\cal K}^{(2)}_{n,n-k-l,k,l}
\epsilon_{n-k-l}^{\bf x}\epsilon_k^{\bf x}\epsilon_l^{\bf x},
\end{equation}
as for instance contributions in $(\epsilon_2^{\bf x})^3$ to $v_2$ or $v_6$~\cite{Borghini:2018xum}.
However, contributions to $v_n(t)$ that are quartic in the eccentricities --- for instance a contribution in $(\epsilon_2^{\bf x})^4$ to $v_8$ --- cannot appear at order $\Kn^{-2}$ (and thus scales as ${\cal O}(t^{n+2})$), but only at order $\Kn^{-3}$ (resp.\ ${\cal O}(t^{n+3})$).

Summarizing, we have found the following scaling behavior of $v_n(t)$ at early times, starting from an initial state without anisotropic flow:
\begin{equation}
\label{vn_vs_t_gen}
v_n(t) \!\underset{\text{early }t}{\sim}\sum_{j\geq 1}\big(\Kn^{-1}\big)^{\!j} \sum_{k\geq 0} a^{(j)}_{n,k} t^{n+2k+j}.
\end{equation}
In the case of a system of particles undergoing elastic binary rescatterings and leaving aside quantum effects from the collision kernel, the coefficient $a^{(j)}_{n,k}$, which differ from one harmonic to the other, may contain products of one, two, till at most $j+1$ spatial eccentricities $\epsilon_l^{\bf x}$, see e.g.\ Eqs.~\eqref{vn_vs_t^(n+1)} (for $j=1$) and \eqref{vn_vs_t^(n+2)} (for $j=2$).

\subsection{Spatial characteristics}
\label{ss:<r^2>,eps^x_coll}

The expansion~\eqref{Taylordg3} also allows one to determine to the early-time development of quantities that characterize the spatial geometry of the system. 
In contrast to the anisotropic flow harmonics, these quantities are usually non-zero in the initial state, and in addition they already evolve in a free-streaming system, i.e.\ under the influence of the first term $\ffs$ in Eq.~\eqref{Taylordg3}.
The subsequent terms, involving the collision kernel, thus describe the change in the early-time development due to rescatterings.
As examples, we shall now discuss the behaviors of the root-mean-square transverse radius $\mean{r_\perp^2}$ or the eccentricities $\epsilon_n^{\bf x}$ defined by Eq.~\eqref{eccentricities}.

In this subsection, we shall make extensive use of the property (valid at any fixed ${\bf x}$)~\cite{DeGroot:1980dk}
\begin{equation}
\label{C[f]_integral}
\int\!\big[a({\bf x}) + p_\mu b^\mu({\bf x})\big]{\cal C}[f]\,
\frac{\dd^2\bf p}{E} = 0,
\end{equation}
for arbitrary functions $a$ and $b^\mu$ (with $\mu\in\{0,1,2\}$) of position. 
Physically, the identity encodes the conservation of particle number, energy and momentum in the binary collisions.

\subsubsection{Root-mean-square radius}

Let us first study how the typical size of the system, which expands into the vacuum, increases. 
For that purpose, we can for instance characterize the system size by the mean-square (transverse) radius
\begin{equation}
\label{r_perp^2(t)}
\mean{r_\perp^2}_t\equiv
\frac{\displaystyle\int\!(x^2+y^2)f(t,{\bf x},{\bf p})\,{\rm d}^2{\bf x}\,{\rm d}^2{\bf p}}%
{\displaystyle\int\!f(t,{\bf x},{\bf p})\,{\rm d}^2{\bf x}\,{\rm d}^2{\bf p}}.
\end{equation}
The idea of the following calculations is readily extended to other powers of $r_\perp$. 
From Sect.~\ref{s:free-stream} we know that the mean-square radius already increases for a free-streaming system, namely as
\begin{equation}
\label{r_perp^2(t)_fs}
\mean{r_\perp^2}_t^{\rm f.s.} = \mean{r_\perp^2}_0 + \mean{\bm{v}_\perp^2}_0 t^2 = 
\mean{r_\perp^2}_0 + t^2,
\end{equation}
where the second identity follows from assuming massless particles propagating in two dimensions only.
Our task is to determine how rescatterings modify this behavior at early times. 

The denominator in Eq.~\eqref{r_perp^2(t)} is the total particle number of the system and remains constant if only elastic rescatterings are present in the system. 
In the following, we thus focus on the numerator of Eq.~\eqref{r_perp^2(t)}. 
Inserting expansion~\eqref{Taylordg3} in this numerator, the linear term in $t$ from the rescatterings involves an integral of the form
\begin{equation}
\label{trick}
\int\!g({\bf x})\,{\cal C}[f]\,{\rm d}^2{\bf x}\,\frac{{\rm d}^2{\bf p}}{E},
\end{equation}
with $g({\bf x}) = x^2 + y^2$, independent of momentum, in the present case.  
The integral over ${\bf p}$ at a fixed position ${\bf x}$ is of the form~\eqref{C[f]_integral} --- with only $a({\bf x}) = g({\bf x})$ being non-zero --- and hence vanishes. 
Thus there is no contribution linear in time to the early time development of $\mean{r_\perp^2}_t$ in the classical Boltzmann scenario, but we can already note that this is not necessarily true for collision kernels that do not implement particle-number conservation. 

Turning next to the contribution from the term due to collisions in $t^2$ in expansion~\eqref{Taylordg3}, it also vanishes for a collision kernel ${\cal C}[f]$ conserving particle number and if the initial distribution is isotropic in momentum space. 
First, one can argue that an integral of the form~\eqref{trick} with any time derivative $\partial_t^k{\cal C}[f]$ instead of ${\cal C}[f]$ is also zero when the collision kernel conserves particle number. 
Indeed, taking the time derivative out of the integral over the phase space variables yields 
\begin{equation}
\label{trick2}
\int\!g({\bf x})\,\partial_t^k{\cal C}[f] \,{\rm d}^2{\bf x}\,\frac{{\rm d}^2{\bf p}}{E} = 
\partial_t^k\!\!\int\!g({\bf x})\,{\cal C}[f] \,{\rm d}^2{\bf x}\,\frac{{\rm d}^2{\bf p}}{E} = 0, 
\end{equation}
since one eventually computes the total time derivative of a function which is identically zero at any time.
This ensures that the term in  $\partial_t{\cal C}[f]\big|_{0\,}t^2$ does not contribute to $\mean{r_\perp^2}_t$ at early times.  

In turn, we argue in Appendix~\ref{app:<r^2>_at_O(t^2)} that the other collision-induced term at order $t^2$ in expansion~\eqref{Taylordg3}, that involving $({\bf p}/E^2)\cdot\bm{\nabla}_{\!x}{\cal C}[f]\big|_0$, does not contribute either to the early-time development of the mean-square radius if the initial distribution is isotropic in momentum.

Thus the elastic binary rescatterings described by the Boltzmann collision kernel do not affect the leading contribution to the increase of the root-mean-square transverse radius $\mean{r_\perp^2}_t$ of the system, which remains dominated by the free-streaming expansion at early times.
The first modification from rescatterings to that behavior can generally occur at order $t^3$ in the expansion~\eqref{Taylordg3}, and thus are subleading compared to the ballistic motion of the particles, at least at early times. 
\begin{equation}
\label{rsquared2}
\mean{r_\perp^2}_t = \mean{r_\perp^2}_t^{\rm f.s.} + {\cal O}(t^3). 
\end{equation}

Let us sketch how things behave at order $t^3$ and higher, and in particular which terms from expansion~\eqref{Taylordg3} can yield a non-zero contribution to $\mean{r_\perp^2}_t$.
First, the term in $\partial_t^{k-1}{\cal C}[f]\big|_0$ at order $t^k$ consistently gives zero for a particle-number conserving collision kernel, thanks to Eq.~\eqref{trick2}. 
The terms in $({\bf p}\cdot\bm{\nabla}_{\!x})^{k}$ with $k\geq 3$ (thus starting at order $t^4$) also yield zero, irrespective of any assumption on the collision kernel apart from its being zero at infinity. 
Indeed, the contributions of such terms can be integrated by parts over $x$ or $y$ twice, to get rid of $r_\perp^2$, leaving the integral over position space of the derivative of a function that vanishes at infinity: for instance
\[
\int\!x^2\frac{(p_x\partial_x)^k}{E^{k+1}} 
{\cal C}[f]\big|_0\,{\rm d}x = 
2\!\int\!\frac{p_x^k\partial_x^{k-2}}{E^{k+1}} {\cal C}[f]\big|_0\,{\rm d}x = 0.
\]
Interestingly, the terms involving the second spatial derivatives of either ${\cal C}[f]\big|_0$ or its time derivatives --- e.g.\ the first term of order $t^3$ or the second of order $t^4$ in Eq.~\eqref{Taylordg3} --- also do not contribute to $\mean{r_\perp^2}_t$ in the special case of massless particles propagating in two dimensions (or at least such that $p_z=0$), as shown in Appendix~\ref{app:<r^2>_at_O(t^3)}. 
But this need not be true more generally.

Thus the only collision-induced terms that can affect the early-time increase of a two-dimensional gas of massless particles undergoing elastic binary collisions are those in expansion~\eqref{Taylordg3} involving a single power of $({\bf p}/E)\cdot\bm{\nabla}_{\!x}$ applied to the time derivatives of ${\cal C}[f]$ at $t=0$. 
For instance, the second term of order $t^3$ or the third term of order $t^4$ in Eq.~\eqref{Taylordg3}.

All in all, if the phase-space distribution is initially isotropic in momentum at each point in the transverse plane and if its subsequent evolution preserves particle number, then rescatterings only affect the growth of the mean square transverse radius $\mean{r_\perp^2}_t$ at order $t^3$ or higher, i.e.\ subleadingly compared to the effect of the free-streaming expansion.
One can readily check that this behavior also holds for any moment $\mean{r_\perp^n}_t$ with arbitrary integer $n$, using the same arguments.

\subsubsection{Eccentricities}

Consider now the spatial eccentricities~\eqref{eccentricities}. 
We have seen in Sect.~\ref{s:free-stream} that they decrease in an collisionless system without initial anisotropic flow, symbolically in $1/[1+{\cal O}(t^2)]$ at early times. 
In the previous subsection, we saw that rescatterings change the behavior of the term in the denominator of Eq.~\eqref{eccentricities} only at order $t^3$ or higher. 
Accordingly, we shall now only investigate the behavior of $\mean{r_\perp^n\ee^{\ii_{}n\theta}}_t$, which remains constant in time for a free-streaming system, see Eq.~\eqref{<r^nexp(intheta)>_fs}.

Repeating the steps used in computing $\mean{r_\perp^2}_t$, one quickly finds that the contributions to $\mean{r_\perp^n\ee^{\ii_{}n\theta}}_t$ in $t$ and $t^2$ vanish.
Again, the term in $t$ and one of the terms in $t^2$ [that with the time derivative of the collision kernel in expansion~\eqref{Taylordg3}] are zero thanks to particle-number conservation --- technically, invoking Eqs.~\eqref{trick}--\eqref{trick2} with $g({\bf x}) = r_\perp^n\ee^{\ii_{}n\theta}$.
The remaining term in $t^2$ is also zero if the initial phase space distribution is isotropic in momentum space, thanks to Eq.~\eqref{int_r^np.nablaC[f]cos(ntheta)} -- which plays here the role played by relation~\eqref{int_r^2p.nablaC[f]} in the calculation of $\mean{r_\perp^2}_t$.

Therefore any change of $\mean{r_\perp^n\ee^{\ii_{}n\theta}}_t$ from rescatterings comes at order $t^3$ or higher. 
Since this is also the order at which rescatterings affect the evolution of $\mean{r_\perp^n}_t$, it will also hold for their ratio, namely the spatial eccentricities~\eqref{eccentricities} [and their generalized version~\eqref{eps_n,m_def}]:
\begin{equation}
\label{eps_vs_t_gen}
\epsilon_n^{\bf x}(t) \!\underset{\text{early }t}{\sim} 
\epsilon_n^{\bf x}(t)\big|^{\rm f.s.} + {\cal O}(t^3),
\end{equation}
where $\epsilon_n^{\bf x}(t)\big|^{\rm f.s.}$ denotes the time dependence of the  eccentricity $\epsilon_n^{\bf x}$ in a free-streaming system with the same initial phase space distribution, as computed in Sect.~\ref{s:free-stream}.
This means that at early times the effect of rescatterings on $\epsilon_n^{\bf x}(t)$ is subleading compared to that of the free-streaming expansion. 

As a final remark, one would intuitively expect a connection between the early time behaviors of $\epsilon_n^{\bf x}(t)$ and $v_n(t)$: 
a non-zero anisotropic flow coefficient $v_n$ clearly affects the spatial eccentricity $\epsilon_n^{\bf x}$, in particular the azimuthally asymmetric numerator $\mean{r_\perp^n\ee^{\ii_{}n\theta}}_t$. 
However, we argued in the previous section and the present one that different terms in Eq.~\eqref{Taylordg3} are responsible for the respective onsets of $v_n(t)$ and $\epsilon_n^{\bf x}(t)$. 
From that observation we would tentatively conclude that in the absence of initial anisotropic flow, the early time behaviors of spatial and momentum anisotropies are not related, or at least not obviously.

\section{Generalizations}
\label{s:generalization}

In this section we discuss how the early-time behavior of the anisotropic flow harmonics [Eq.~\eqref{vn_vs_t_gen}] and the spatial eccentricities [Eq.~\eqref{eps_vs_t_gen}] change when one departs from the two-dimensionless system of massless ``classical'' particles with elastic binary rescatterings assumed in Sect.~\ref{s:interacting-system2}.

\subsection{3-dimensional expansion}
\label{ss:3D}

What happens if the system expands in 3 dimensions instead of 2 in the previous section? 
Clearly, this should slow down the transverse expansion, since $\mean{v_\perp^2}$ is now in general smaller than 1, so that the characteristic time scale $\sqrt{\mean{r_\perp^2}/\mean{v_\perp^2}}$ increases (by a factor $\sqrt{3/2}$, in case the velocity distribution is also isotropic along the third direction).
But the actual question is, whether the scaling behaviors found in Sect.~\ref{s:interacting-system2} are modified or not. 

Looking back at the reason why $v_n(t)$ grows in $t^{n+1}$ at early times in the absence of initial anisotropic flow, we see that the arguments that were used never involved the dimensionality of the space into which the system is expanding. 
Instead, the proof rather relied on the necessity to have a term that contains $n$ powers of the transverse momentum ${\bf p}$ and that is not odd in any of the other momenta appearing in the collision kernel.\footnote{In particular, the longitudinal components of the momenta play no role in the reasoning. Accordingly, whether the longitudinal motion of the system is boost invariant or not does not affect the scaling behaviors.}
That these requirements are not fulfilled at any order $t^k$ with $k\leq n$ remains true in the case of a three-dimensional expansion, and thus the scaling behavior~\eqref{vn_vs_t_gen}, including the dependence on the inverse Knudsen number, still holds. 
Of course the coefficients $a^{(j)}_{n,k}$ do depend on whether the system is expanding in two or three dimensions. 
But their generic dependence on the initial spatial eccentricities is unchanged: for instance, the $a^{(1)}_{n,k}$, which are the relevant coefficients at order $\Kn^{-1}$, only depend linearly or quadratically on the initial $\{\epsilon_l^{\bf x}\}$.

Similarly, the space dimension did not play any role in our reasoning at orders ${\cal O}(t)$ and ${\cal O}(t^2)$ in the calculation of the mean square radius (or more generally $\mean{r_\perp^n}_t$) and the spatial eccentricities $\epsilon_n^{\bf x}$. 
We actually encountered a term that vanishes only for the special case of massless particles propagating in two dimensions (or with $p_z = 0$), but only at order ${\cal O}(t^3)$, i.e.\ subleading with respect to the free-streaming behavior. 
So here again, the scaling law~\eqref{eps_vs_t_gen} determined in Sect.~\ref{ss:<r^2>,eps^x_coll} is still valid in the three-dimensional case.

\subsection{Massive particles}
\label{ss:mass}

Another assumption of Sect.~\ref{s:interacting-system2} was that of considering massless degrees of freedom. 
This hypothesis leads to a simpler expression of the M{\o}ller velocity~\eqref{vrel} [see Eq.~\eqref{vrel_2D_m=0}], which is helpful when dealing with explicit semi-analytical examples~\cite{Borghini:2022qha}.
However this feature is actually irrelevant for our derivations of the scaling behavior of anisotropic flow and the eccentricities. 

Indeed, the M{\o}ller velocity of massive particles with momenta $\vecp$, $\vecp_1$ can only depend on $\vecp$ via the inner product $\vecp\cdot\vecp_1$,\footnote{The square modulus $\vecp^2$ ``does not know'' about the azimuth $\varphi_{\bf p}$, and thus cannot contribute to anisotropic flow.} just like in the massless case.
To contribute to the early-time development of a flow harmonic $v_n(t)$, a factor $(\vecp\cdot\vecp_1)^k$ from expanding $\vrel$ has to multiply a contribution $(\vecp_1\cdot\vec{\nabla}_{\!x})^k$ to match the $k$ powers of $\vecp_1$ from the M{\o}ller velocity, and another contribution $(\vecp\cdot\vec{\nabla}_{\!x})^{n-k}$, to obtain the necessary $n$ powers of $\vecp$. 
That is, $n$ powers of the spatial gradient $\vec{\nabla}_{\!x}$ should appear in the relevant term in expansion~\eqref{Taylordg3}, and this is only possible at order ${\cal O}(t^{n+1})$ or higher. 
Accordingly, the scaling behavior~\eqref{vn_vs_t_gen} remains valid in the case of massive particles. 

It is also clear that the early-time behavior~\eqref{eps_vs_t_gen} of the spatial eccentricities remains valid too if the particles are massive. 
Indeed, the arguments to establish that the possible contributions from scatterings at orders ${\cal O}(t)$ and ${\cal O}(t^2)$ are zero were very general ones (particle number conservation, momentum isotropy of the initial distribution) and totally irrespective of any detail of the collision kernel --- apart from its conserving particle number.

Let us also mention in this subsection another possible modification of the composition of the system that does not affect the scaling behaviors~\eqref{vn_vs_t_gen} and \eqref{eps_vs_t_gen}.
Until now we only considered systems consisting of a single species of particles. 
In heavy ion collisions, it is certainly more realistic to consider a mixture of several species that can not only self-interact but also interact with each other. 
One should then introduce several phase space distributions, which obey coupled Boltzmann equations~\cite{DeGroot:1980dk}.
If the number of particles of each species remains constant, i.e.\ if all possible scattering processes are elastic, then one can check that the scaling behavior~\eqref{vn_vs_t_gen} of the anisotropic-flow coefficients is unchanged, since the arguments used to derive it still hold. 
One then finds that the flow coefficients $v_n(t)$ of a given species depend not only on the initial eccentricities of that species itself, but also on those of the other species, which may be interesting if different particle species have different initial geometrical profiles due to different production mechanisms~\cite{Kersting:2018qvi}.

\subsection{Inclusion of quantum statistics in the collision kernel}
\label{ss:quantum_C[f]}

In this subsection and the the next we investigate how modifying the form of the collision kernel ${\cal C}[f]$ affects the scaling behaviors of the flow harmonics $v_n(t)$ and the eccentricities $\epsilon_n^{\bf x}$ of Sect.~\ref{s:interacting-system2}.

The first modification we consider consists in incorporating quantum effects --- Pauli blocking or Bose enhancement --- into the binary collision kernel~\eqref{C[f]_2->2}--\eqref{C[f]_2->2_sym}, via the usual substitution 
\begin{equation}
f'f'_1 - ff_1 \to f'f'_1FF_1 - ff_1F'F'_1
\label{quantum_C[f]}
\end{equation}
in the integrand, where $F(t,\vec{x},\vecp)\equiv 1\pm f(t,\vec{x},\vecp)$.

Now, the extra terms introduced in the collision kernel by the substitution~\eqref{quantum_C[f]} do not introduce any extra power of $\vecp$ or the spatial gradient $\vec{\nabla}_{\!x}$, and thus they cannot modify the scaling behavior~\eqref{vn_vs_t_gen} of the flow harmonics. 
Similarly, the modified collision term still conserves particle number, and also leaves the early-time behavior~\eqref{eps_vs_t_gen} qualitatively unchanged. 

However, as was already noted in Ref.~\cite{Borghini:2018xum} the longer form of the collision kernel due to the change~\eqref{quantum_C[f]} does induce a modification, namely in the dependence of the coefficients $a^{(j)}_{n,k}$ in Eq.~\eqref{vn_vs_t_gen} on the initial spatial eccentricities $\{\epsilon_l^{\bf x}\}$. 
Indeed, since products of three or four distribution functions now appear in the integrand of ${\cal C}[f]$, the coefficients $a^{(1)}_{n,k}$ can now include terms depending cubically or quartically on the initial eccentricities. 
For instance, the early-time expansion $v_2(t)$ or $v_6(t)$ can now have a term in $(\epsilon_2^{\bf x})^3$ at leading order $\Kn^{-1}$, or $v_4(t)$ a contribution in $(\epsilon_2^{\bf x})^4$. 
In practice, this will only be the case if the initial state is dense enough for quantum corrections to become relevant.

\subsection{Alternative collision kernel}
\label{ss:other_C[f]}

Another type of modification of the collision kernel ${\cal C}[f]$ has a higher impact on the scaling behaviors of Sect.~\ref{s:interacting-system2}, namely if we drop the assumption that the rescatterings in the system are elastic. 
Instead, one can include inelastic two-to-two ($2\leftrightarrow 2$) processes (if the system consists of several species), as well as particle-number changing processes: 
for instance $1\leftrightarrow 2$ gluon splitting or fusion processes~\cite{Geiger:1991nj,Arnold:2002zm} or $2\leftrightarrow 3$ parton processes~\cite{Xu:2004mz}.

Indeed, in Sect.~\ref{ss:<r^2>,eps^x_coll} we invoked several times particle-number conservation to cancel terms when deriving the scaling behaviors of the mean square radius $\mean{r_\perp^2}_t$ [Eq.~\eqref{rsquared2}] or of $\mean{r_\perp^n\ee^{\ii_{}n\theta}}_t$, and as a result of the spatial eccentricities, Eq.~\eqref{eps_vs_t_gen}. 
If particle number is not conserved, then the early-time evolution of every average $\mean{r_\perp^n}_t$ or $\mean{r_\perp^n\ee^{\ii_{}n\theta}}_t$ and therefore of $\epsilon_n^{\bf x}(t)$ will generally include linear and quadratic terms in $t$, both at order $\Kn^{-1}$, which represents a significant modification of the free-streaming evolution. 

Turning to the anisotropic flow coefficients $v_n(t)$, their early-time behavior is also affected if the collision kernel does not conserve particle number, although much less than the spatial eccentricities. 
As a matter of fact, particle-number conservation plays no role in the time dependence of the numerator of Eq.~\eqref{vn_vs_f(t,x,p)}, which determined the overall scaling of $v_n(t)$ in Sect.~\ref{ss:vn}. 
On the other hand, the denominator of Eq.~\eqref{vn_vs_f(t,x,p)} is obviously no longer constant if particle number is not conserved, which has to be taken into account. 
By integrating expansion~\eqref{Taylordg3} first over $\vec{x}$ (so that total-gradient terms vanish) and then over $\vecp$ (where only terms even in all momenta can contribute), one finds using the same arguments as in Sect.~\ref{ss:vn} that the total particle number behaves at early times like
\begin{equation}
N(t) \!\underset{\text{early }t}{\sim} 
N_0 + \sum_{j\geq 1}\big(\Kn^{-1}\big)^{\!j}\sum_{k\geq 0}b_k^{(j)}t^{2k+j},
\end{equation}
where $N_0$ denotes its initial value. 
Combining this behavior with that of the numerator and reorganizing the double expansion in powers of $\Kn^{-1}$ and $t$, one finds that $v_n(t)$ still obeys a scaling law of the form~\eqref{vn_vs_t_gen}, with modified coefficients $a^{(j)}_{n,k}$ --- apart from the leading coefficient $a^{(1)}_{n,0}$ of the term in $\Kn^{-1}t^{n+1}$, which is the same as if $N(t)$ stayed constant.

\subsection{Initial anisotropic flow}
\label{ss:initial_vn}

Eventually, a last key ingredient in Sect.~\ref{s:interacting-system2} was the hypothesis of vanishing initial anisotropic flow. 
This assumption may naturally be released, not only out of mathematical curiosity, but also for several physically motivated reasons: 
If $t_0$ still represents the initial time of the fireball expansion, one may consider that anisotropic flow is already present due to initial-state correlations, as investigated for instance in AMPT in Ref.~\cite{Zhang:2015cya}, or that the finite particle multiplicity unavoidably leads to (small) flow harmonics. 
Or $t_0$ may instead represent a later time in the system evolution, say the time at which the description changes from fluid dynamics to a transport approach, in which case the presence of anisotropic flow results from the preceding evolution.  

In any case, if the assumption of initial momentum isotropy is released, then one should beware that the free-streaming behaviors are modified. 
Indeed, in the absence of rescatterings each coefficient $v_n(t)$ remains at its initial value, whether it vanishes or not. 
In turn, the spatial eccentricities no longer decrease as found in Sect.~\ref{s:free-stream}. 
For instance, one can see than an initial $v_n$ will spoil relation~\eqref{<r^nexp(intheta)>_fs} and lead to a time-dependent average $\mean{r_\perp\ee^{\ii_{}n\theta}}_t^{\rm f.s.}$ instead.

That being told, the scaling behaviors~\eqref{vn_vs_t_gen} and \eqref{eps_vs_t_gen}, now viewed as describing the departure from the free-streaming behaviors, will generally no longer hold.
Thus we have seen in Sect.~\ref{ss:vn} that the assumption of vanishing initial anisotropic flow was necessary to cancel the influence on the flow harmonics of the term linear in $t$ in expansion~\eqref{Taylordg3}. 
If there is some initial anisotropic flow $v_n(t_0)$, then $v_n(t)$ may depart linearly from that value at (shortly) later times $t$, as was found on a toy example in Ref.~\cite{Borghini:2011qc}.
Similarly, in Sect.~\ref{ss:<r^2>,eps^x_coll} initial momentum isotropy was a necessary ingredient to cancel one of the terms in $t^2$ contributing to the mean square radius $\mean{r_\perp^2}_t$ or the asymmetry $\mean{r_\perp^2\ee^{\ii_{}n\theta}}_t$, so that departure from momentum isotropy will generally lead to terms at order ${\cal O}(t^2)$ in the scaling behavior~\eqref{eps_vs_t_gen}.

\section{Discussion}
\label{s:discussion}

The main result of this paper is that in a system described by the kinetic Boltzmann equation with a particle-number conserving collision kernel, and with transverse momentum isotropy in the initial state, the anisotropic flow coefficients $v_n(t)$ resp.\ the spatial eccentricities $\epsilon_n^{\bf x}(t)$ scale according to Eq.~\eqref{vn_vs_t_gen} resp.\ Eq.~\eqref{eps_vs_t_gen} at early times. 
In the case of the harmonics $v_n$, we also detailed the dependence on the average number of rescatterings per particle, characterized here by the inverse Knudsen number $\Kn^{-1}$.

These behaviors are quite generic, since they follow from writing the early-time expansion of the phase space distribution as a Taylor series~\eqref{Taylordg3} and invoking general principles. 
They thus hold whether the system is expanding in two or three dimensions, and whether it consists of massless or massive classical or quantum particles. 
Simplifying the collision kernel, for instance using the popular relaxation time approximation~\cite{Kurkela:2018ygx,Kurkela:2020wwb,Rocha:2021zcw,Ambrus:2021fej}, should also not spoil the scaling behaviors~\eqref{vn_vs_t_gen}, \eqref{eps_vs_t_gen}, at least as long as conservation laws are properly implemented.
The only exceptions we encountered are twofold, namely if there is already anisotropic flow in the initial state or if the particles can undergo inelastic collisions. 
Actually, the latter possibility only affects the scaling of eccentricities, not that of the flow harmonics, whose behavior is thus more robust. 

The early-time scaling $v_n(t) \propto t^{n+1}$ of momentum anisotropies in transport calculations was already observed in special cases before: 
in numerical simulations with a given initial phase distribution in Ref.~\cite{Alver:2010dn}, and in analytical calculations restricted to leading order in $\Kn^{-1}$ with another specific initial profile~\cite{Borghini:2010hy}. 
Here we have shown that this scaling behavior is very general if there is no initial flow. 

The most studied anisotropic-flow harmonic at ultrarelativistic energies, both experimentally and theoretically, is elliptic flow, for which Eq.~\eqref{vn_vs_t_gen} yields
\begin{equation}
\label{v2_vs_t^3}
v_2(t) \!\underset{\text{early }t}{\propto}\! \Kn^{-1\,} t^3,
\end{equation} 
at lowest order in $t$. 
This behavior should be contrasted with the findings in fluid dynamics, either in simulations~\cite{Heinz:2002rs,Kolb:2002cq,Kolb:2003dz,Teaney:2010vd} or using general arguments~\cite{Vredevoogd:2008id}, which give $v_2(t) \propto t^2$ at early times.\footnote{Strictly speaking, the measure of the momentum anisotropy is sometimes $v_2$ itself, and sometimes related to the asymmetry $T^{xx}-T^{yy}$ of the energy-momentum tensor, which necessitates no particlization of the fluid. The latter also scales as $t^3$ for the system investigated in the present paper~\cite{Borghini:2022qha}.}
The mismatch between the scaling exponents predicted by kinetic theory and fluid dynamics was already noted in Ref.~\cite{Alver:2010dn} but only for two-dimensional expansions. 
In this paper we showed that the difference remains for kinetic theory in three dimensions. 
This implies that there is no ``universal behavior'' for the development of momentum anisotropies --- in contrast to that found for radial transverse flow~\cite{Vredevoogd:2008id} --- but rather that different classes of theories (kinetic theory or fluid dynamics) may lead to different behaviors. 
If there exist ``late-time attractor solutions'' for truly three-dimensional dynamical scenarios without cylindrical symmetry, as was found empirically for one-dimensional motion in both strong and weak coupling regimes~\cite{Heller:2015dha,Romatschke:2017vte,Kurkela:2019set,Denicol:2019lio}, then it would be interesting to see how this could be reconciled with the different behaviors of anisotropic flow at early times. 

The results reported in the present paper are admittedly rather formal, because we deliberately aimed at staying as general as possible, with minimal assumptions on the initial phase-space distribution and the collision kernel. 
This is what allows us to derive generic results. 
In a companion paper~\cite{Borghini:2022qha} we shall present comparisons with numerical simulations, including a few straightforward generalizations --- for instance, looking at energy-weighted anisotropies. 
Accordingly, we shall be able to illustrate the effect of including higher-order terms in $t$ and/or $\Kn^{-1}$ from expansion~\eqref{Taylordg3}. 
The price to pay is that we use a single, simple setup (two-dimensional system of massless particles, toy initial profile) and thus lose the generality of the present paper.

Eventually, one should naturally ask whether the different scaling behaviors of say $v_2(t)$ in kinetic theory vs.\ fluid dynamics are relevant for heavy-ion phenomenology.
The main issue is clearly that experimentally only the final values of the flow harmonics, and to a lesser extent the spatial eccentricities, are directly accessible.
In collisions of heavy nuclei, one may hope that accessing some characteristics of the early-time dynamics of the bulk could become feasible by studying particles that decouple early from the system, as e.g.\ photons or dileptons pairs in an appropriate invariant-mass interval~\cite{Coquet:2021lca}. 
The onset of ``pre-flow'' anisotropies of the bulk in large systems may also be relevant for determining the initial condition to the subsequent fluid-dynamical evolution. 
Accessing the early-time behavior may be more feasible in ``small systems'', since in that case the evolution lasts less long, so that the final flow values are possibly more influenced by the early stage.
One should however remember that the characteristic time scale of the early-time evolution also becomes smaller in such systems.

\begin{acknowledgments}
We thank Nina Kersting and Hendrik Roch for discussions. 
We acknowledge support by the Deutsche Forschungsgemeinschaft (DFG, German Research Foundation) through the CRC-TR 211 ``Strong-interaction matter under extreme conditions'' -- project number 315477589 -- TRR 211.
\end{acknowledgments}

\appendix

\section{Time derivative of the collision kernel: an example}
\label{app:d_tC[f]}

To illustrate the procedure for eliminating the time derivatives of the collision term ${\cal C}[f]$, we take as an example the kernel~\eqref{C[f]_2->2_sym}, which we recast in the shorter form
\begin{equation}
\label{C[f]_2->2_sym_simple}
{\cal C}[f] = \frac{1}{2}\!\int_{{\bf p}_1,{\bf p}',{\bf p}'_1}\!
(f'f'_1 - ff_1)_{}W({\bf p},{\bf p}_1\to{\bf p}',{\bf p}'_1),
\end{equation}
in which the precise form of the integration measure in momentum space is no longer written. 
Differentiating this kernel with respect to time yields at once
\begin{widetext}
\begin{equation}
\partial_t{\cal C}[f] = \frac{1}{2}\!\int_{{\bf p}_1,{\bf p}',{\bf p}'_1}\!
\big[(\partial_t f')f'_1 + (\partial_t f'_1)f' 
- (\partial_t f)f_1 - (\partial_t f_1)f\big]  
W({\bf p},{\bf p}_1\to{\bf p}',{\bf p}'_1).
\end{equation}
If we now invoke the Boltzmann equation~\eqref{Boltzmann-eq}, we can re-express every time derivative of the phase space distribution in the integrand, which leads to
\begin{align}
\partial_t{\cal C}[f] = &\ \frac{1}{2}\!\int_{{\bf p}_1,{\bf p}',{\bf p}'_1}\!
\bigg(\!\!-\!\frac{{\bf p}'\cdot\bm{\nabla}_{\!x} f'}{E'}f'_1 - 
\frac{{\bf p}'_1\cdot\bm{\nabla}_{\!x} f'_1}{E'_1}f' + 
\frac{{\bf p}\cdot\bm{\nabla}_{\!x} f}{E}f_1 + 
\frac{{\bf p}_1\cdot\bm{\nabla}_{\!x} f_1}{E_1}f\bigg)
W({\bf p},{\bf p}_1\to{\bf p}',{\bf p}'_1) \cr
&+ \frac{1}{2}\!\int_{{\bf p}_1,{\bf p}',{\bf p}'_1}\!
\bigg(\frac{{\cal C}[f']}{E'}_{}f'_1 + \frac{{\cal C}[f'_1]}{E'_1}_{}f' - 
\frac{{\cal C}[f]}{E}_{}f_1 + \frac{{\cal C}[f_1]}{E_1}_{}f \bigg)
W({\bf p},{\bf p}_1\to{\bf p}',{\bf p}'_1).
\label{d_tC[f]_example}
\end{align}
The term on the right hand side of the first line is of order $\Kn^{-1}$, while that in the second line is of order $\Kn^{-2}$, as mentioned in Sec.~\ref{s:interacting-system1}.

Iterating the procedure, one finds that the $k$-th time derivative $\partial_t^k{\cal C}[f]$ will involve the $k$-th power of the spatial gradient, $(\bm{\nabla}_{\!x})^k$, at order $\Kn^{-1}$, together with lower powers $(\bm{\nabla}_{\!x})^{k-j}$ at order $\big(\Kn^{-1}\big)^{j+1}$ for all $1\leq j\leq k$.
For instance, the second time derivative reads
\begin{align}
\partial_t^2{\cal C}[f] = &\ \frac{1}{2}\int_{{\bf p}_1,{\bf p}',{\bf p}'_1}\!
\bigg[\frac{\big({\bf p}'\cdot\bm{\nabla}_{\!x}\big)^{\!2\!} f'}{E^{\prime 2}}f'_1 + 
\frac{\big({\bf p}'_1\cdot\bm{\nabla}_{\!x}\big)^{\!2\!} f'_1}{E^{\prime 2}_1}f' - 
\frac{\big({\bf p}\cdot\bm{\nabla}_{\!x}\big)^{\!2\!} f}{E^2}f_1 - 
\frac{\big({\bf p}_1\cdot\bm{\nabla}_{\!x}\big)^{\!2\!} f_1}{E_1^2}f\bigg]
W({\bf p},{\bf p}_1\to{\bf p}',{\bf p}'_1)\quad \cr
&+ \!\int_{{\bf p}_1,{\bf p}',{\bf p}'_1}\!
\bigg(\frac{{\bf p}'\cdot\bm{\nabla}_{\!x} f'}{E'}
\frac{{\bf p}'_1\cdot\bm{\nabla}_{\!x} f'_1}{E'_1} - 
\frac{{\bf p}\cdot\bm{\nabla}_{\!x} f}{E}
\frac{{\bf p}_1\cdot\bm{\nabla}_{\!x} f_1}{E_1} \bigg)
W({\bf p},{\bf p}_1\to{\bf p}',{\bf p}'_1) + {\cal O}(\Kn^{-2},\Kn^{-3}),
\label{d_t^2C[f]_example}
\end{align}
where we only wrote the term of (leading) order $\Kn^{-1}$. 
The term of order $\Kn^{-2}$ is
\begin{align}
&\int_{{\bf p}_1,{\bf p}',{\bf p}'_1}\!
\bigg(\!\!-\!\frac{{\bf p}'\cdot\bm{\nabla}_{\!x} f'}{E'}\frac{{\cal C}[f'_1]}{E'_1} - 
\frac{{\bf p}'_1\cdot\bm{\nabla}_{\!x} f'_1}{E'_1}\frac{{\cal C}[f']}{E'} + 
\frac{{\bf p}\cdot\bm{\nabla}_{\!x} f}{E}\frac{{\cal C}[f_1]}{E_1} + 
\frac{{\bf p}_1\cdot\bm{\nabla}_{\!x} f_1}{E_1}\frac{{\cal C}[f]}{E}\bigg)
W({\bf p},{\bf p}_1\to{\bf p}',{\bf p}'_1) \cr
&\ + \int_{{\bf p}_1,{\bf p}',{\bf p}'_1}\!
\bigg(\!\!-\!\frac{{\bf p}'\cdot\bm{\nabla}_{\!x}{\cal C}[f']}{E^{\prime 2}}f'_1 - 
\frac{{\bf p}'_1\cdot\bm{\nabla}_{\!x}{\cal C}[f'_1]}{E^{\prime 2}_1}f' + 
\frac{{\bf p}\cdot\bm{\nabla}_{\!x}{\cal C}[f]}{E^2}f_1 + 
\frac{{\bf p}_1\cdot\bm{\nabla}_{\!x}{\cal C}[f_1]}{E_1^2}f\bigg)
W({\bf p},{\bf p}_1\to{\bf p}',{\bf p}'_1),
\label{d_t^2C[f]_example2}
\end{align}
\end{widetext}
while that of order $\Kn^{-3}$ includes in its integrand contributions of the type ${\cal C}[f]_{}{\cal C}[f_1]$ and ${\cal C}[{\cal C}[f]]$, both multiplying the transition rate $W$.

\section{Calculation of contributions to the early-time development of spatial characteristics}
\label{app:<r^n>}

\subsection{Contributions at order ${\cal O}(t^2)$}
\label{app:<r^2>_at_O(t^2)}

Let us show that the integral
\begin{equation}
\label{int_r^2p.nablaC[f]}
\int\!(x^2+y^2)\frac{{\bf p}\cdot\bm{\nabla}_{\!x}}{E^2}_{} {\cal C}[f]\big|_0\,
{\rm d}^2{\bf p}\,{\rm d}^2{\bf x},
\end{equation}
which enters the early-time behavior of the mean square radius $\mean{r_\perp^2}_t$ at order $t^2$, vanishes when the initial distribution is isotropic in momentum. 

Performing integration by parts over the spatial variables using the fact that ${\cal C}[f]\big|_0$ vanishes at large distances, the integral becomes
\[
-2\!\int\!\frac{xp_x + yp_y}{E^2}_{} {\cal C}[f]\big|_0\,
{\rm d}^2{\bf p}\,{\rm d}^2{\bf x}.
\]
The integral over momentum is then vanishing at every ${\bf x}$:
Replacing ${\cal C}[f]$ by its explicit expression, one obtains an integral over all momenta of an integrand which is always odd in ${\bf p}$ or ${\bf p}_1$ (when the M{\o}ller velocity is involved), since $f_0$ is isotropic in momentum space.

The same reasoning gives
\begin{equation}
\label{int_r^np.nablaC[f]}
\int\!(x^2+y^2)^{n/2\,}\frac{{\bf p}\cdot\bm{\nabla}_{\!x}}{E^2}_{} {\cal C}[f]\big|_0\,
{\rm d}^2{\bf p}\,{\rm d}^2{\bf x} = 0,
\end{equation}
resp.
\begin{equation}
\label{int_r^np.nablaC[f]cos(ntheta)}
\int\!r_\perp^n\ee^{\ii_{}n\theta\,}\frac{{\bf p}\cdot\bm{\nabla}_{\!x}}{E^2}_{} {\cal C}[f]\big|_0\,
{\rm d}^2{\bf p}\,{\rm d}^2{\bf x} = 0,
\end{equation}
which is relevant for the average $\mean{r_\perp^n}_t$ resp.\ $\mean{r_\perp^n\ee^{\ii_{}n\theta}}_t$ at order $t^2$.

\subsection{Influence of mass on the early-time development of $\mean{r_\perp^2}_t$}
\label{app:<r^2>_at_O(t^3)}

In this appendix we detail how the particle mass affects the integral
\begin{equation}
\label{integral_nr.N}
\int\!(x^2+y^2)\frac{\big({\bf p}\cdot\bm{\nabla}_{\!x}\big)^2}{E^3} {\cal C}[f]\big|_0\,
{\rm d}^2{\bf p}\,{\rm d}^2{\bf x},
\end{equation}
which appears when one investigates the behavior of the mean-square radius $\mean{r_\perp^2}_t$ at order $t^3$ (see Sec.~\ref{ss:<r^2>,eps^x_coll}).

Writing $({\bf p}\cdot\bm{\nabla}_{\!x})^2 = p_x^2\partial_x^2 + 2p_xp_y\partial_x\partial_y + p_y^2\partial_y^2$, one can first handle the term in $p_x^2\partial_x^2$ by performing two successive integrations by parts over $x$. 
Since ${\cal C}[f]\big|_0$ or its derivative $\partial_x{\cal C}[f]\big|_0$ vanish as $|x|\to\infty$ --- the system has a finite initial size ---, one finds
\[
\int\!(x^2+y^2)\frac{p_x^2}{E^3}\partial_x^2 {\cal C}[f]\big|_0\,{\rm d}x = 
2\frac{p_x^2}{E^3}\!\int\!{\cal C}[f]\big|_0\,{\rm d}x.
\]
With the help of the similar result for the term in $p_y^2\partial_y^2$, the contribution from the two corresponding terms to the integral~\eqref{integral_nr.N} becomes
\begin{equation}
\label{integral_nr.N+1}
2\!\int\!\frac{p_x^2+p_y^2}{E^2}_{} {\cal C}[f]\big|_0\,
\frac{{\rm d}^2{\bf p}}{E}\,{\rm d}^2{\bf x}.
\end{equation}
If the particles are massless, the ratio in the integrand equals 1, and the integral over ${\bf p}$ vanishes for a particle-number-conserving collision kernel, thanks to property~\eqref{C[f]_integral}. 

Eventually, the mixed term in $\partial_x\partial_y$ does not contribute to the integral~\eqref{integral_nr.N} either. 
Indeed, the integral over the spatial variables are readily performed and yield ${\cal C}[f]\big|_0$ (or one of its spatial derivatives) at infinity, where it vanishes. 

Altogether, we have thus showed that the integral~\eqref{integral_nr.N} vanishes for a two-dimensional system of massless particles.
This is no longer necessarily true for massive particles, or if they propagate in three dimensions.

\end{document}